\documentstyle[11pt,psfig,twoside]{article}

\begin{document}

\newcommand{\dd}{\,{\rm d}}
\newcommand{\ie}{{\it i.e.},\,}
\newcommand{\etal}{{\it et al.\ }}
\newcommand{\eg}{{\it e.g.},\,}
\newcommand{\cf}{{\it cf.\ }}
\newcommand{\vs}{{\it vs.\ }}
\newcommand{\zdot}{\makebox[0pt][l]{.}}
\newcommand{\up}[1]{\ifmmode^{\rm #1}\else$^{\rm #1}$\fi}
\newcommand{\dn}[1]{\ifmmode_{\rm #1}\else$_{\rm #1}$\fi}
\newcommand{\upd}{\up{d}}
\newcommand{\uph}{\up{h}}
\newcommand{\upm}{\up{m}}
\newcommand{\ups}{\up{s}}
\newcommand{\arcd}{\ifmmode^{\circ}\else$^{\circ}$\fi}
\newcommand{\arcm}{\ifmmode{'}\else$'$\fi}
\newcommand{\arcs}{\ifmmode{''}\else$''$\fi}
\newcommand{\MS}{{\rm M}\ifmmode_{\odot}\else$_{\odot}$\fi}
\newcommand{\RS}{{\rm R}\ifmmode_{\odot}\else$_{\odot}$\fi}
\newcommand{\LS}{{\rm L}\ifmmode_{\odot}\else$_{\odot}$\fi}

\newcommand{\Abstract}[2]{{\footnotesize\begin{center}ABSTRACT\end{center}
\vspace{1mm}\par#1\par
\noindent
{~}{\it #2}}}

\newcommand{\TabCap}[2]{\begin{center}\parbox[t]{#1}{\begin{center}
  \small {\spaceskip 2pt plus 1pt minus 1pt T a b l e}
  \refstepcounter{table}\thetable \\[2mm]
  \footnotesize #2 \end{center}}\end{center}}

\newcommand{\TableSep}[2]{\begin{table}[p]\vspace{#1}
\TabCap{#2}\end{table}}

\newcommand{\FigCap}[1]{\footnotesize\par\noindent Fig.\  %
  \refstepcounter{figure}\thefigure. #1\par}

\newcommand{\TableFont}{\footnotesize}
\newcommand{\TableFontIt}{\ttit}
\newcommand{\SetTableFont}[1]{\renewcommand{\TableFont}{#1}}

\newcommand{\MakeTable}[4]{\begin{table}[htb]\TabCap{#2}{#3}
  \begin{center} \TableFont \begin{tabular}{#1} #4 
  \end{tabular}\end{center}\end{table}}

\newcommand{\MakeTableSep}[4]{\begin{table}[p]\TabCap{#2}{#3}
  \begin{center} \TableFont \begin{tabular}{#1} #4 
  \end{tabular}\end{center}\end{table}}

\newenvironment{references}%
{
\footnotesize \frenchspacing
\renewcommand{\thesection}{}
\renewcommand{\in}{{\rm in }}
\renewcommand{\AA}{Astron.\ Astrophys.}
\newcommand{\AAS}{Astron.~Astrophys.~Suppl.~Ser.}
\newcommand{\ApJ}{Astrophys.\ J.}
\newcommand{\ApJS}{Astrophys.\ J.~Suppl.~Ser.}
\newcommand{\ApJL}{Astrophys.\ J.~Letters}
\newcommand{\AJ}{Astron.\ J.}
\newcommand{\IBVS}{IBVS}
\newcommand{\PASP}{P.A.S.P.}
\newcommand{\Acta}{Acta Astron.}
\newcommand{\MNRAS}{MNRAS}
\renewcommand{\and}{{\rm and }}
\section{{\rm REFERENCES}}
\sloppy \hyphenpenalty10000
\begin{list}{}{\leftmargin1cm\listparindent-1cm
\itemindent\listparindent\parsep0pt\itemsep0pt}}%
{\end{list}\vspace{2mm}}

\def\TYLDA{~}
\newlength{\DW}
\settowidth{\DW}{0}
\newcommand{\dw}{\hspace{\DW}}

\newcommand{\refitem}[5]{\item[]{#1} #2%
\def\REFARG{#3}\ifx\REFARG\TYLDA\else, {\it#3}\fi
\def\REFARG{#4}\ifx\REFARG\TYLDA\else, {\bf#4}\fi
\def\REFARG{#5}\ifx\REFARG\TYLDA\else, {#5}\fi.}

\newcommand{\Section}[1]{\section{#1}}
\newcommand{\Subsection}[1]{\subsection{#1}}
\newcommand{\Acknow}[1]{\par\vspace{5mm}{\bf Acknowledgements.} #1}
\pagestyle{myheadings}

\def\thefootnote{\fnsymbol{footnote}}
\begin{center}
{\Large\bf The Optical Gravitational Lensing Experiment.\\
\vskip3pt
Catalog of Star Clusters from the Large Magellanic Cloud\footnote{Based
on observations obtained with the 1.3~m Warsaw telescope at 
the Las Campanas Observatory operated by the Carnegie Institution of 
Washington.}} 

\vskip 1cm

{ G.~~P~i~e~t~r~z~y~{\'n}~s~k~i$^1$,~~A.~~U~d~a~l~s~k~i$^1$,~~M.~~K~u~b~i~a~k$^1$,\\
M.~~S~z~y~m~a~{\'n}~s~k~i$^1$,~~P.~~W~o~{\'z}~n~i~a~k$^2$,~~and~~K.~~{\.Z}~e~b~r~u~{\'n}$^1$}

\vskip5mm

{$^1$Warsaw University Observatory, Al.~Ujazdowskie~4, 00-478~Warszawa,
Poland\\
e-mail: (pietrzyn,udalski,mk,msz,zebrun)@astrouw.edu.pl\\
$^2$ Princeton University Observatory, Princeton, NJ 08544-1001, USA\\ 
e-mail: wozniak@astro.princeton.edu }

\end{center}

\Abstract{We present the catalog of star clusters found in the area of
about  5.8 square degree in the central regions of the Large Magellanic
Cloud. It   contains data for 745 clusters. 126 of them are new objects.
For each cluster  equatorial coordinates, radius, approximate number of
members and cross-identification are provided. Photometric data for all 
clusters presented in the catalog and Atlas consisting of finding charts
and color-magnitude diagrams are available electronically from the OGLE
Internet  archive.}{~} 

\Section{Introduction}
The Optical Gravitational Lensing Experiment (OGLE) is a long term observing 
project with the main goal to provide information on dark unseen matter using 
microlensing events (Paczy{\'n}ski 1986). Detailed description of the project 
can be found in Udalski, Kubiak and Szyma{\'n}ski (1997). 

The main observing targets of the second phase of the survey, OGLE-II,
include large parts of the central bars of the SMC and LMC. Photometry
collected in the standard {\it BVI} bands for  millions of stars located
in dense and  poorly observed so far central regions  of these galaxies
provides an ideal material for many side projects.  

One of the sub-projects of the OGLE-II survey aims at searching for and 
analyzing properties of star clusters in the Magellanic Clouds. In 
Pietrzy{\'n}ski \etal (1998, hereafter Paper~I) the catalog of 273 clusters 
from the central parts of the SMC  was published. In this paper we present 
the catalog of clusters from the observed regions in the LMC. 

OGLE catalogs of star clusters enable  many further detailed studies  of
properties of star clusters in the Magellanic Clouds. Results concerning
 the system of star clusters in the SMC were already published. Ages of
93 objects from the catalog of clusters from the SMC were derived using
standard  procedure of isochrone fitting by Pietrzy{\'n}ski and Udalski
(1999a). The  multiple cluster candidates were selected and listed in
Pietrzy{\'n}ski and  Udalski (1999b). Similar data for clusters from the
LMC will be published in  the forthcoming papers. 

Large number observations collected in the course of the OGLE-II survey
provides  also an unique opportunity to explore populations of variable
stars in star  clusters. In the first paper of the series  on variable
stars located in the  regions of star clusters 127 eclipsing systems in
optical coincidence with  star clusters in the SMC were presented by
Pietrzy{\'n}ski and Udalski  (1999c). The lists of Cepheids located in
the close neighborhood of the Magellanic Cloud clusters were published by
Pietrzy{\'n}ski and Udalski (1999d). Further papers  containing lists as
well as more detailed studies of variable stars from  clusters of the
Magellanic Clouds observed by the OGLE collaboration will  follow.

Bearing in mind the potential usefulness of the photometric data of
presented  star clusters we have decided to make them publicly available. They
are accessible from  the OGLE Internet archive. 

\Section{Previous Searches} Many efforts have been done for searching
for star clusters in the LMC since  Shapley and Mohr (1932) identified
clusters in this galaxy for the first time.  The large atlas of clusters
covering most of the LMC was published by Hodge  and Wright (1967). This
catalog contains 1146 objects discovered until 1967,  by Lynga and
Westerlund (1963), Shapley and Lindsay (1963) and  Hodge and  Sexton
(1966). Lamberts (1982) cataloged objects based on the ESO survey. This 
atlas provided coordinates for many star clusters, however majority of
them  were already known. Olszewski \etal (1988) found additional 156
clusters outside the area covered by the Hodge and Wright catalog. Hodge
(1980, 1988) performed  deep search for clusters in the LMC using 4-m
telescope plates. He reported  387 new clusters in 15 small regions.
Kontizas \etal (1990) constructed a new  catalog of clusters based on
the ESO/SERC Southern Sky Atlas. They presented  1762 clusters in the
large ${25\times25}$ degree area centered on the LMC but  excluding the
crowded regions around the bar. About 600 of them were new  objects. 
   
All this searches were made visually. The first automatic search for
clusters  in the LMC was performed by Bhatia and MacGillivray (1989).
This project  resulted in detection of 284 objects in a 6 square degree
field, which  quadrupled the number of previously known clusters in this
region. Zaritsky,  Harris and Thompson (1997) presented {\it UBVI} CCD
photometry of stars from  ${2\times1.5}$ degree field located northwest
of the LMC bar. Based on this  data they found 68 clusters (about 45\%
more than previously detected).  Recently Bica \etal (1999) published a
comprehensive catalog of extended objects  from the LMC. 

As can be seen from this short review many  potential  clusters have
been  discovered in the LMC so far. It should be stressed, however, that
some of these  detections may turn out to be spurious. For example
Kontizas \etal (1990) did   not confirm 210 objects from the previous
catalogs. The automatic techniques appear to be very effective in
searching for star clusters in the LMC.  Unfortunately they were applied
to relatively small regions located outside  the crowded area near the
central bar. 

Although the number of known clusters in the LMC is large, deep and precise 
photometry exists only for very limited sample of the populous LMC clusters.  
The aim of this paper is to provide the astronomical community with precise 
photometry of  relatively large sample of clusters from the LMC, selected in 
the algorithmic way.

\Section{Observations}
All observations presented in this paper were collected during the second 
phase of the OGLE microlensing survey with the 1.3~m Warsaw telescope at the 
Las Campanas Observatory, Chile, which is operated by the Carnegie Institution 
of Washington. The telescope was equipped with ${2048\times2048}$ CCD detector 
working in driftscan mode. The gain and readout noise were 3.8~e$^-$/ADU 
and 5.4~e$^-$, respectively. Details of the system can be found in Udalski, 
Kubiak and Szyma{\'n}ski (1997). 

\MakeTableSep{lcc}{12.5cm}{Equatorial coordinates of the OGLE-II LMC fields}
{
\hline
\noalign{\vskip3pt}
\multicolumn{1}{c}{Field} & RA (J2000)  & DEC (J2000)\\
\hline
\noalign{\vskip3pt}
LMC$\_$SC1  &  5\uph33\upm49\ups & $-70\arcd06\arcm10\arcs$ \\
LMC$\_$SC2  &  5\uph31\upm17\ups & $-69\arcd51\arcm55\arcs$ \\
LMC$\_$SC3  &  5\uph28\upm48\ups & $-69\arcd48\arcm05\arcs$ \\
LMC$\_$SC4  &  5\uph26\upm18\ups & $-69\arcd48\arcm05\arcs$ \\
LMC$\_$SC5  &  5\uph23\upm48\ups & $-69\arcd41\arcm05\arcs$ \\
LMC$\_$SC6  &  5\uph21\upm18\ups & $-69\arcd37\arcm10\arcs$ \\
LMC$\_$SC7  &  5\uph18\upm48\ups & $-69\arcd24\arcm10\arcs$ \\
LMC$\_$SC8  &  5\uph16\upm18\ups & $-69\arcd19\arcm15\arcs$ \\
LMC$\_$SC9  &  5\uph13\upm48\ups & $-69\arcd14\arcm05\arcs$ \\
LMC$\_$SC10 &  5\uph11\upm16\ups & $-69\arcd09\arcm15\arcs$ \\
LMC$\_$SC11 &  5\uph08\upm41\ups & $-69\arcd10\arcm05\arcs$ \\
LMC$\_$SC12 &  5\uph06\upm16\ups & $-69\arcd38\arcm20\arcs$ \\
LMC$\_$SC13 &  5\uph06\upm14\ups & $-68\arcd43\arcm30\arcs$ \\
LMC$\_$SC14 &  5\uph03\upm49\ups & $-69\arcd04\arcm45\arcs$ \\
LMC$\_$SC15 &  5\uph01\upm17\ups & $-69\arcd04\arcm45\arcs$ \\
LMC$\_$SC16 &  5\uph36\upm18\ups & $-70\arcd09\arcm40\arcs$ \\
LMC$\_$SC17 &  5\uph38\upm48\ups & $-70\arcd16\arcm45\arcs$ \\
LMC$\_$SC18 &  5\uph41\upm18\ups & $-70\arcd24\arcm50\arcs$ \\
LMC$\_$SC19 &  5\uph43\upm48\ups & $-70\arcd34\arcm45\arcs$ \\
LMC$\_$SC20 &  5\uph46\upm18\ups & $-70\arcd44\arcm50\arcs$ \\
LMC$\_$SC21 &  5\uph21\upm14\ups & $-70\arcd33\arcm20\arcs$ \\
LMC$\_$SC22 &  5\uph02\upm26\ups & $-67\arcd09\arcm35\arcs$ \\
LMC$\_$SC23 &  5\uph04\upm45\ups & $-67\arcd09\arcm40\arcs$ \\
LMC$\_$SC24 &  5\uph07\upm05\ups & $-67\arcd09\arcm35\arcs$ \\
LMC$\_$SC25 &  5\uph09\upm24\ups & $-67\arcd09\arcm30\arcs$ \\
LMC$\_$SC26 &  5\uph11\upm43\ups & $-67\arcd09\arcm40\arcs$ \\
\hline}

Observations were conducted in 26 slightly overlapping fields with the size of 
about ${14\zdot\arcm2\times57\arcm}$ each, which gave the total coverage  of 
about 5.8 square degree. Table~1 lists acronyms of the observed 
fields with equatorial coordinates of their centers. Collected images were 
reduced with the standard OGLE data pipeline. Accuracy of transformations to 
the standard system  was $0.01-0.02$ mag. For more details about data reduction 
and transformation procedures the reader is referred to the paper with description 
of {\it BVI} photometric maps of the SMC (Udalski \etal 1998). Quality of the 
data collected for the LMC is similar to that from the SMC. {\it BVI} maps of 
the LMC  will be released in the near future (Udalski \etal in preparation). 

\Section{Catalog}
\Subsection{Search for Clusters}
Visual searches are subjective and, especially in crowded stellar background 
may lead to many spurious detections. The observed area in the LMC is 
relatively dense and large. In order to obtain the objective list of cluster 
candidates we performed an automatic, algorithmic search. Similar technique as 
presented in Zaritsky, Harris and Thompson (1997) was applied. This algorithm 
was already successfully used by us in searching for clusters in the SMC 
(Paper~I). Detailed description of this algorithm can be found in Paper~I. 
In short, each of 26 driftscans was divided into square boxes and stars were 
counted inside each of them. Because of the different size of the potential 
clusters, three sets of such density maps with boxes of ${10\times10}$, 
${20\times20}$ and ${30\times30}$ pixels (${4.1\times4.1}$, ${8.2\times8.2}$, 
${12.3\times12.3}$, arcsec respectively) were constructed. In order to remove 
background variations from our density maps the "unsharp masking" procedure 
was used. The suspected clusters were selected as the concentration of at 
least four pixels above a given threshold from the unsharp masked images. 
Three different detection thresholds of 4, 3 and 2$\sigma$ of all pixels of 
the map were used. 

Then, all candidates were carefully examined and many of them were rejected 
due to proximity to bright overexposed stars or the edge of the frame. Our 
procedure resulted in detection of 745 objects. 

The catalog was compared with the list of clusters presented by Bica \etal 
(1999). 619 objects turned out to be common to both these catalogs. Table~2 
contains description of acronyms used in our catalog. 
\MakeTable{|llr|}{8cm}{LMC catalogs}{
\hline Acronym & Reference & Entries \\ 
\hline H      & Hodge 1960       &   2 \\ 
SL     & Shapley and Lindsay 1963   & 145 \\ 
HS     & Hodge and Sexton    1966   & 108 \\ 
H80    & Hodge               1980   &   2 \\ 
KMK88  & Kontizas \etal 1988       &  44 \\ 
H88    & Hodge 1988                 &  94 \\ 
KMHK   & Kontizas \etal 1990       &  74 \\ 
BRHT   & Bhatia et al. 1991         &  26 \\ 
BCD    & Bica \etal 1992           &   2 \\ 
BCDSP  & Bica \etal 1996           &   1 \\ 
ZHT    & Zaritsky \etal 1997       &   2 \\ 
BSDL   &  Bica \etal 1999          & 221 \\ 
OGLE   & this paper                 & 126 \\ 
\hline} 

\Subsection{Coordinates and Angular Sizes of Clusters}
The equatorial coordinates of clusters were obtained in the identical manner 
as in Paper~I. Their accuracy depends on cluster richness and ranges from 
2~arcsec for compact populous clusters to 10 arcsec for loose  faint ones. 

Angular sizes of clusters were derived based on density profiles
obtained from  star counts performed in consecutive annuli around their
adopted centers. In most cases  counts obtained in relatively large
distance from the cluster  centers allowed to reliably define stellar
background and derive precise  dimensions. Many profiles show, however,
significant fluctuations after the  main drop of the  stellar density.
Such a behavior may be caused by presence  of extended halos around
clusters or background density fluctuations. As in  the case of Paper~I
we decided to define two different kinds of cluster radii.  One accounts
for the presence of the above-mentioned fluctuations and represents  the
radius of entire cluster. The second one is defined as the distance from
the cluster center to the main drop of the stellar density and it is
useful  for defining  the boundary between the regions where the
cluster or field stars are more frequent. 

\begin{figure}[p] \vspace*{-6.3cm}
\centerline{\hspace*{10mm}\psfig{figure=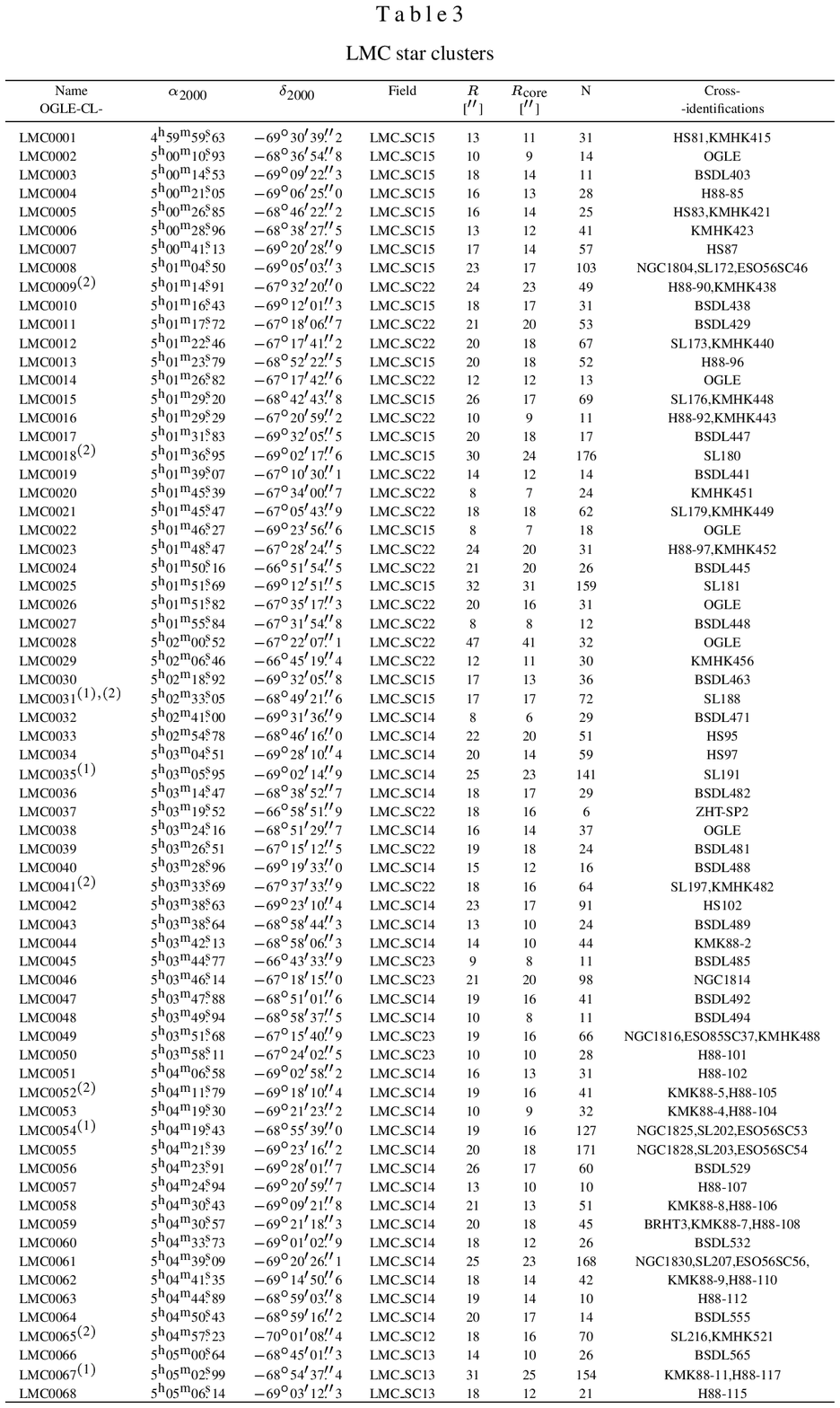,clip=}} \end{figure}
\begin{figure}[p] \vspace*{-6.3cm}
\centerline{\hspace*{10mm}\psfig{figure=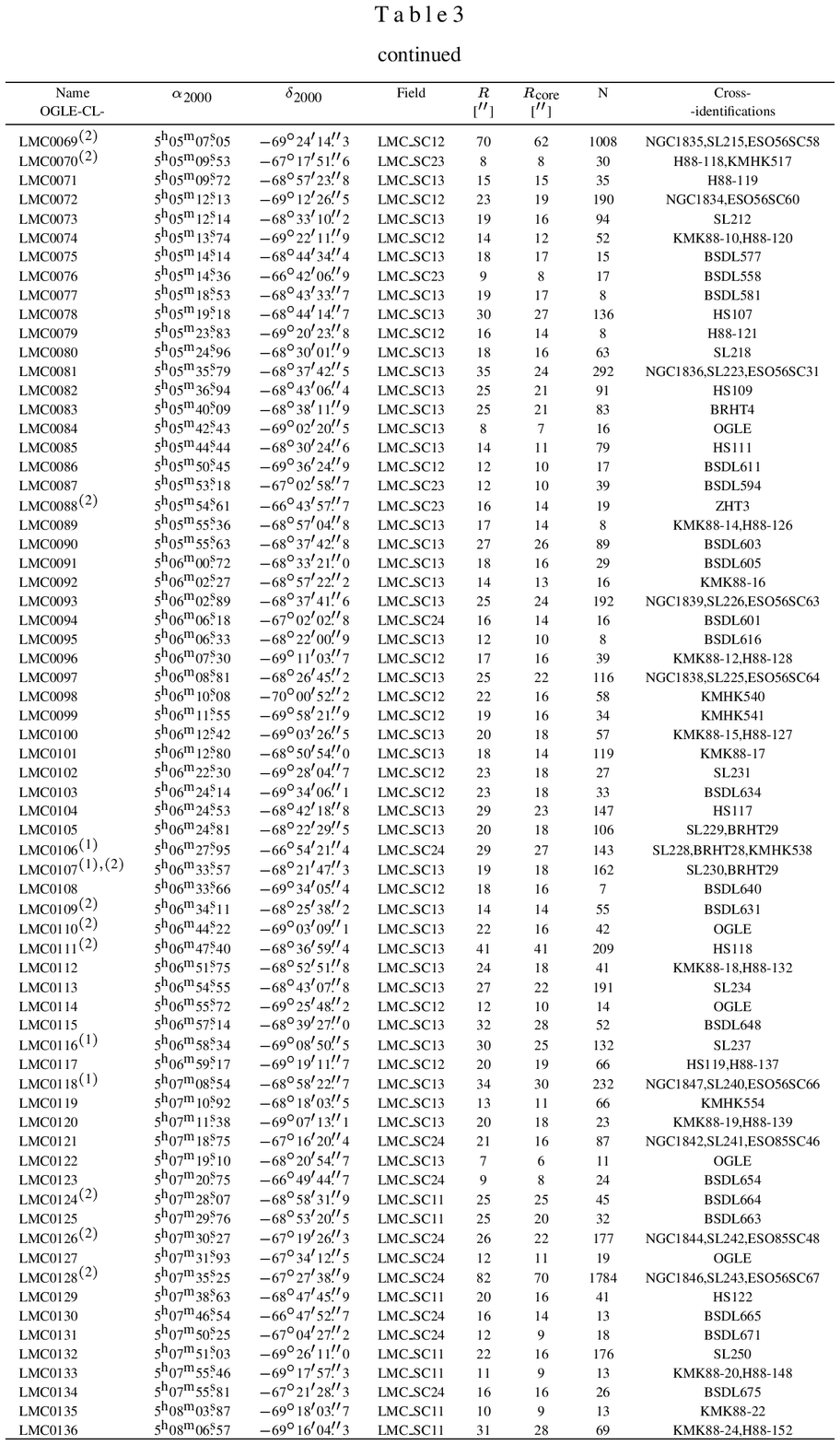,clip=}} \end{figure}
\begin{figure}[p] \vspace*{-6.3cm}
\centerline{\hspace*{10mm}\psfig{figure=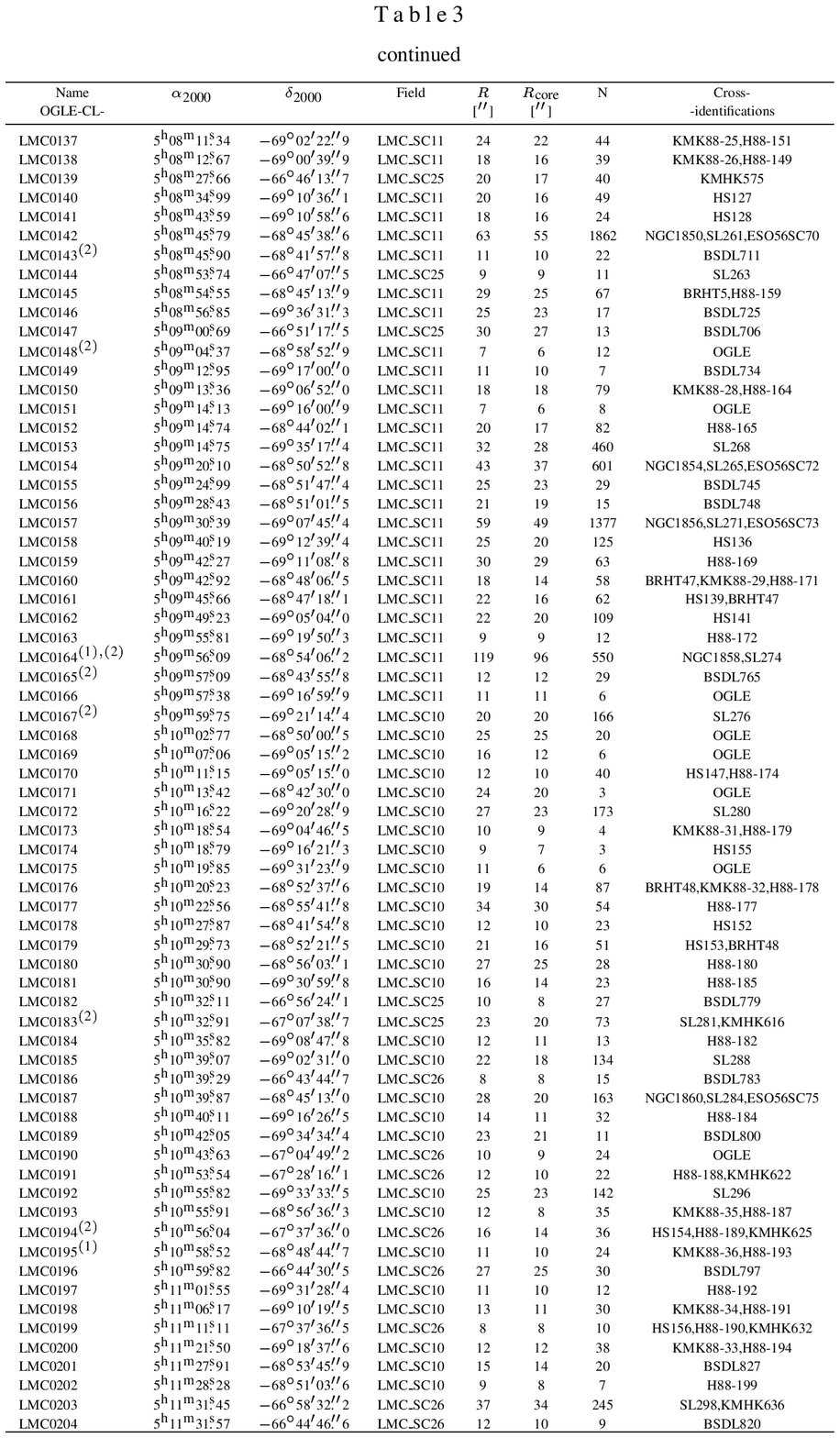,clip=}} \end{figure}
\begin{figure}[p] \vspace*{-6.3cm}
\centerline{\hspace*{10mm}\psfig{figure=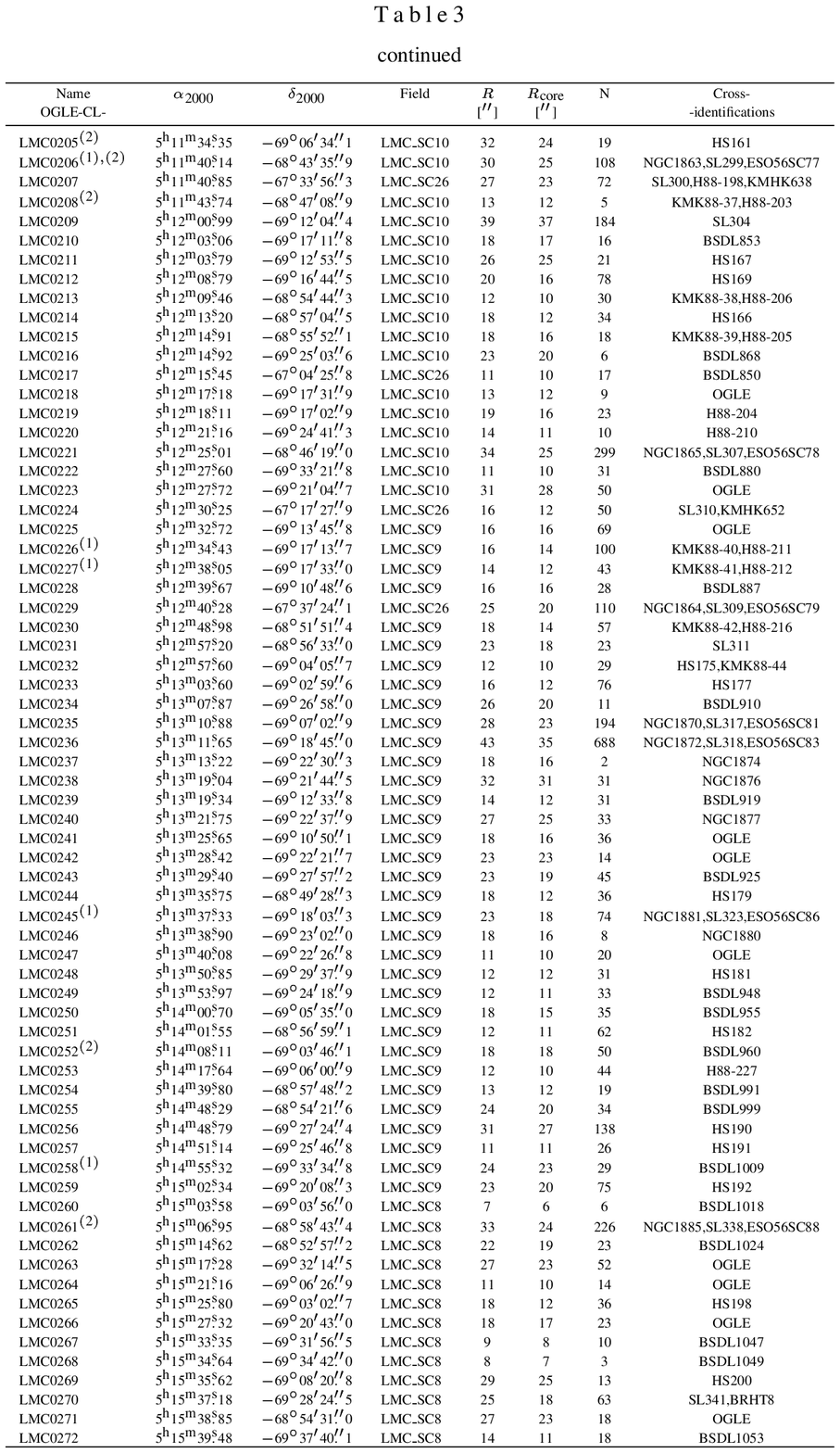,clip=}} \end{figure}
\begin{figure}[p] \vspace*{-6.3cm}
\centerline{\hspace*{10mm}\psfig{figure=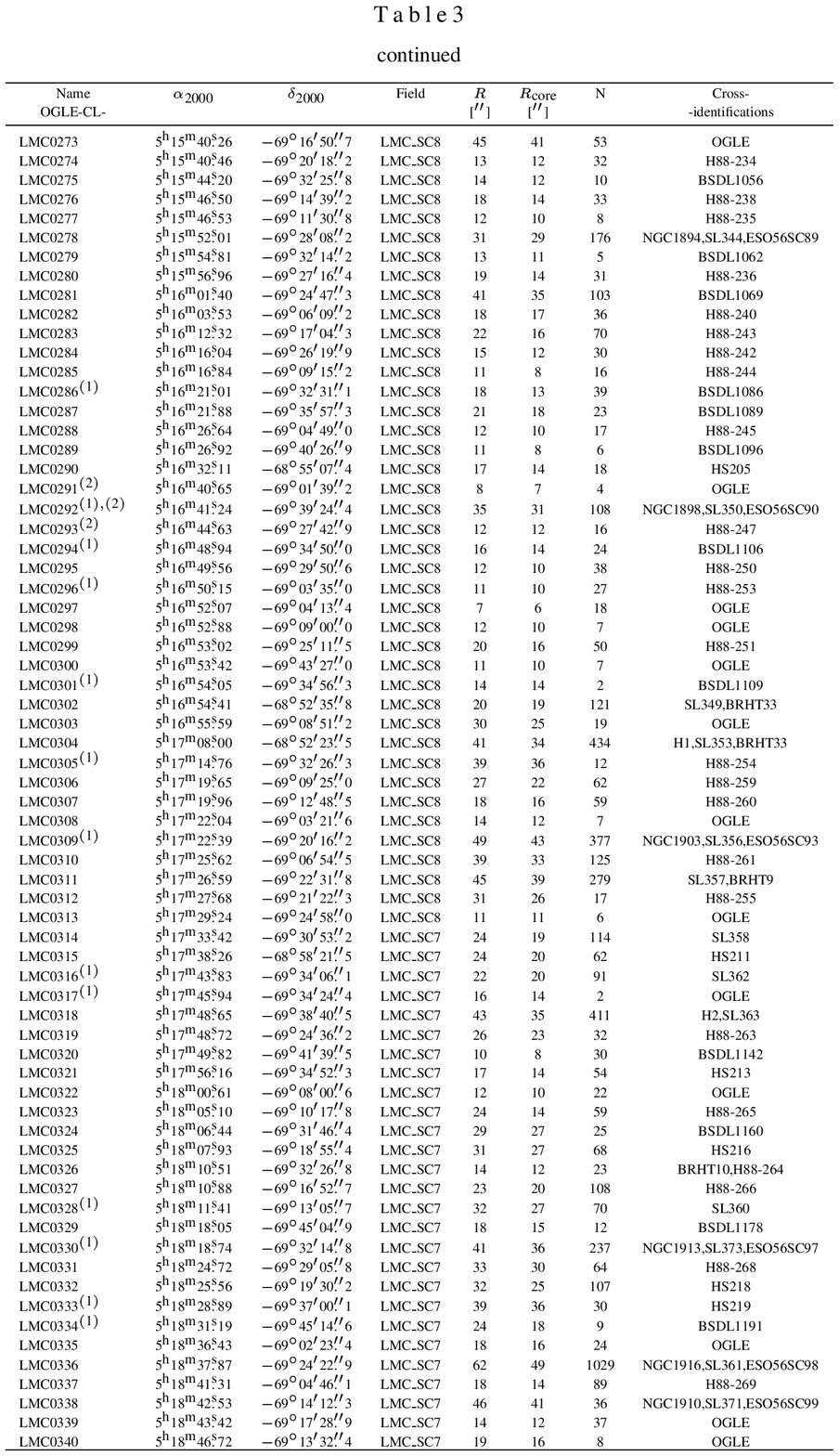,clip=}} \end{figure}
\begin{figure}[p] \vspace*{-6.3cm}
\centerline{\hspace*{10mm}\psfig{figure=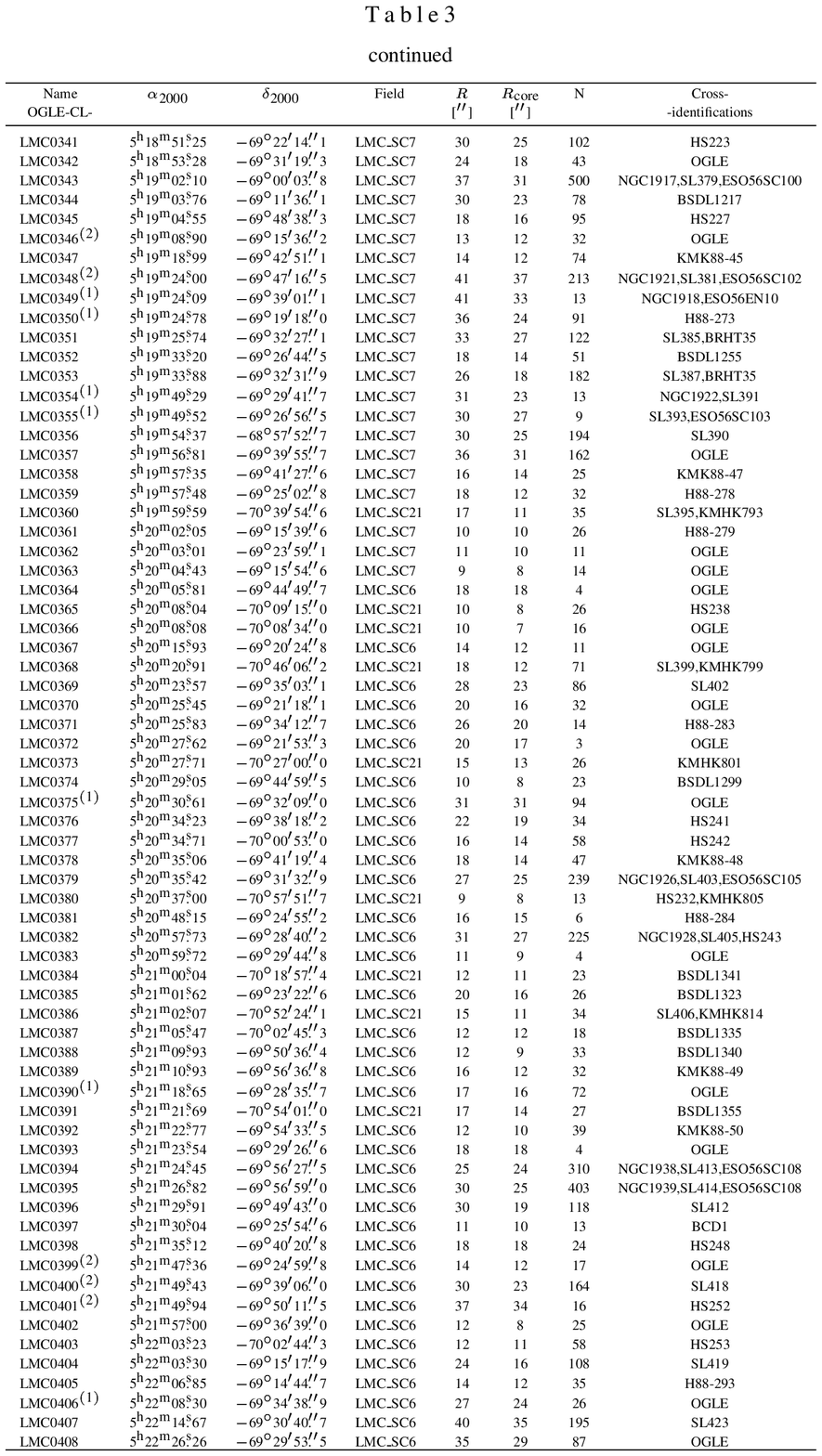,clip=}} \end{figure}
\begin{figure}[p] \vspace*{-6.3cm}
\centerline{\hspace*{10mm}\psfig{figure=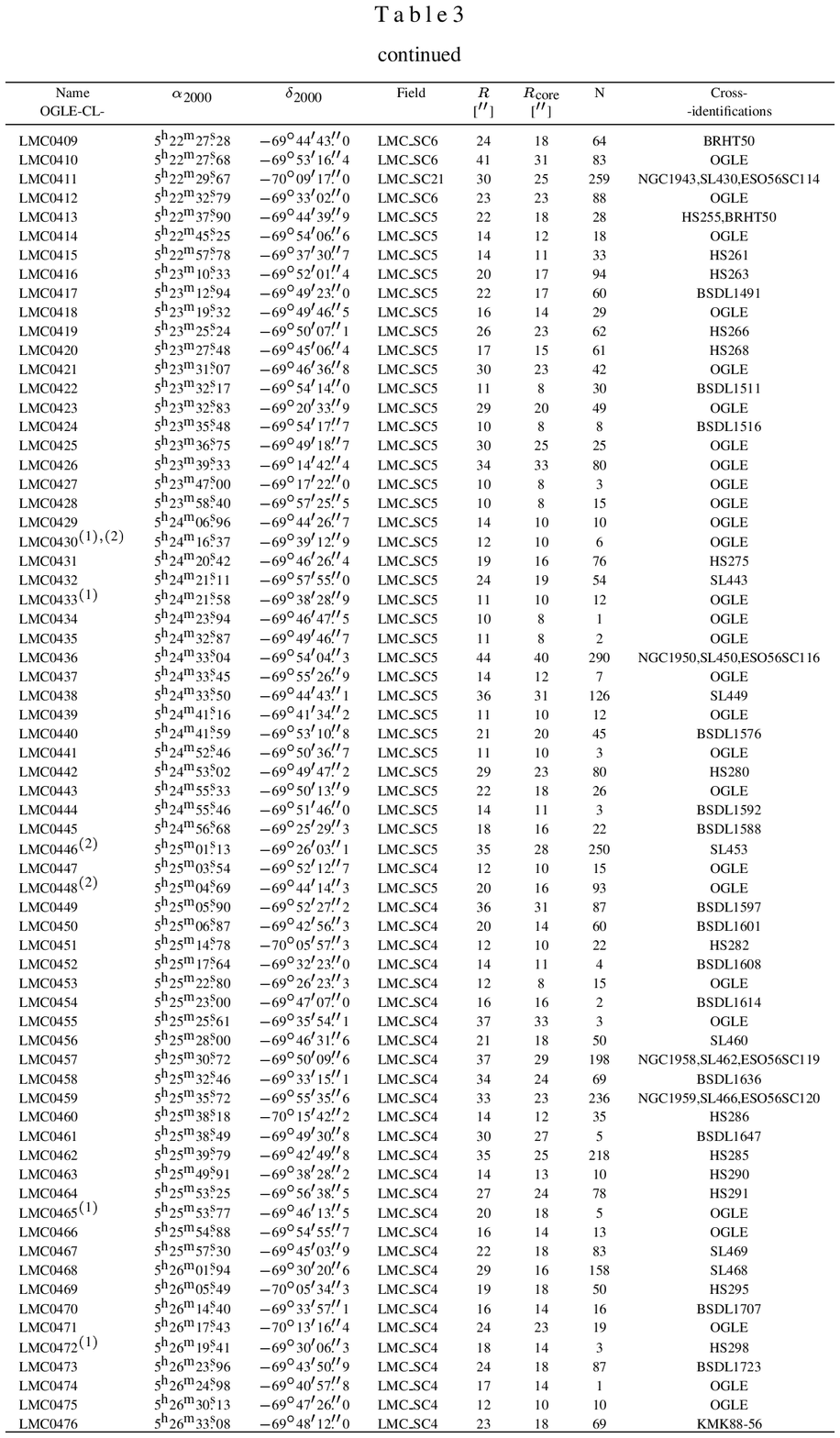,clip=}} \end{figure}
\begin{figure}[p] \vspace*{-6.3cm}
\centerline{\hspace*{10mm}\psfig{figure=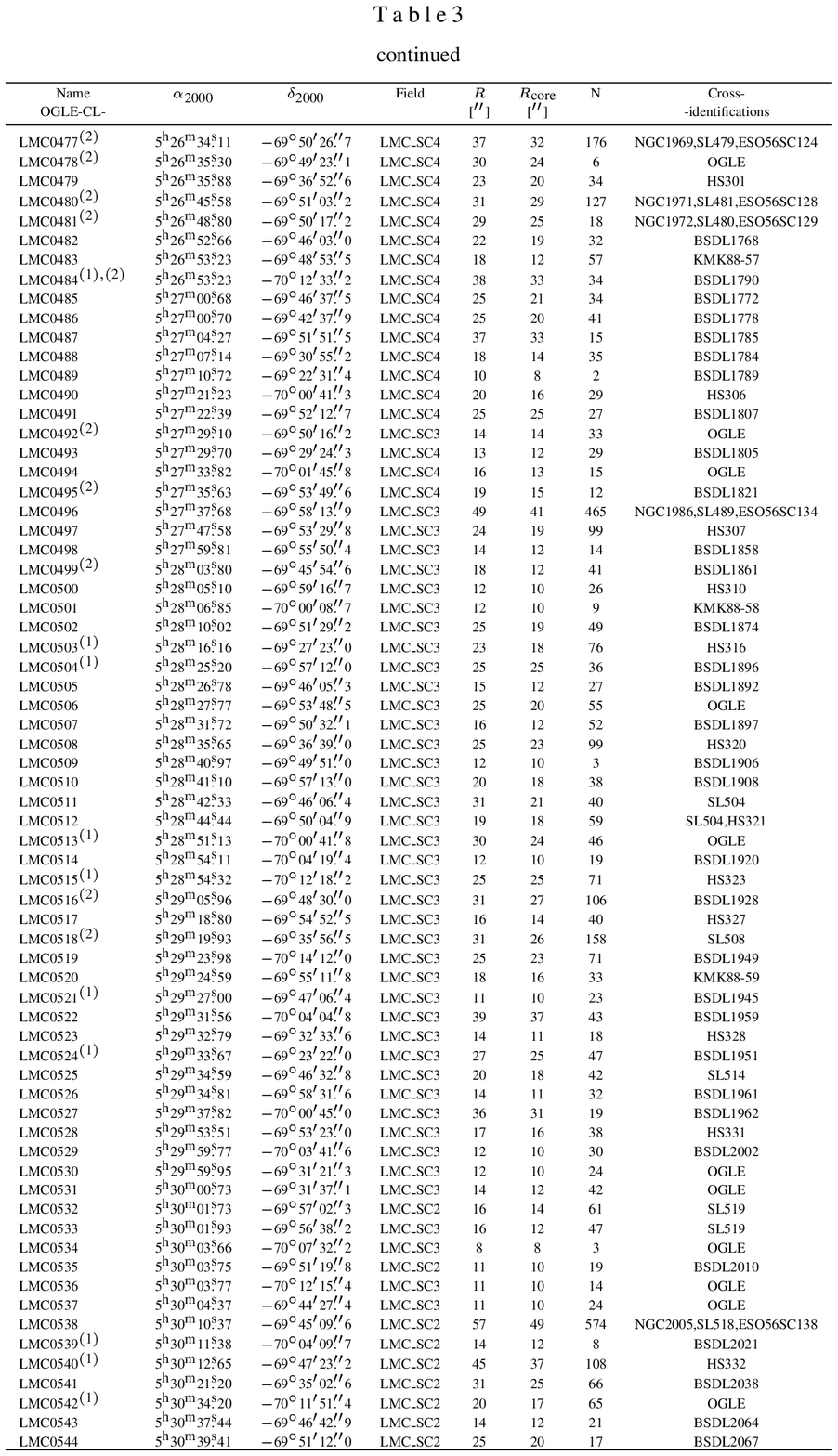,clip=}} \end{figure}
\begin{figure}[p] \vspace*{-6.3cm}
\centerline{\hspace*{10mm}\psfig{figure=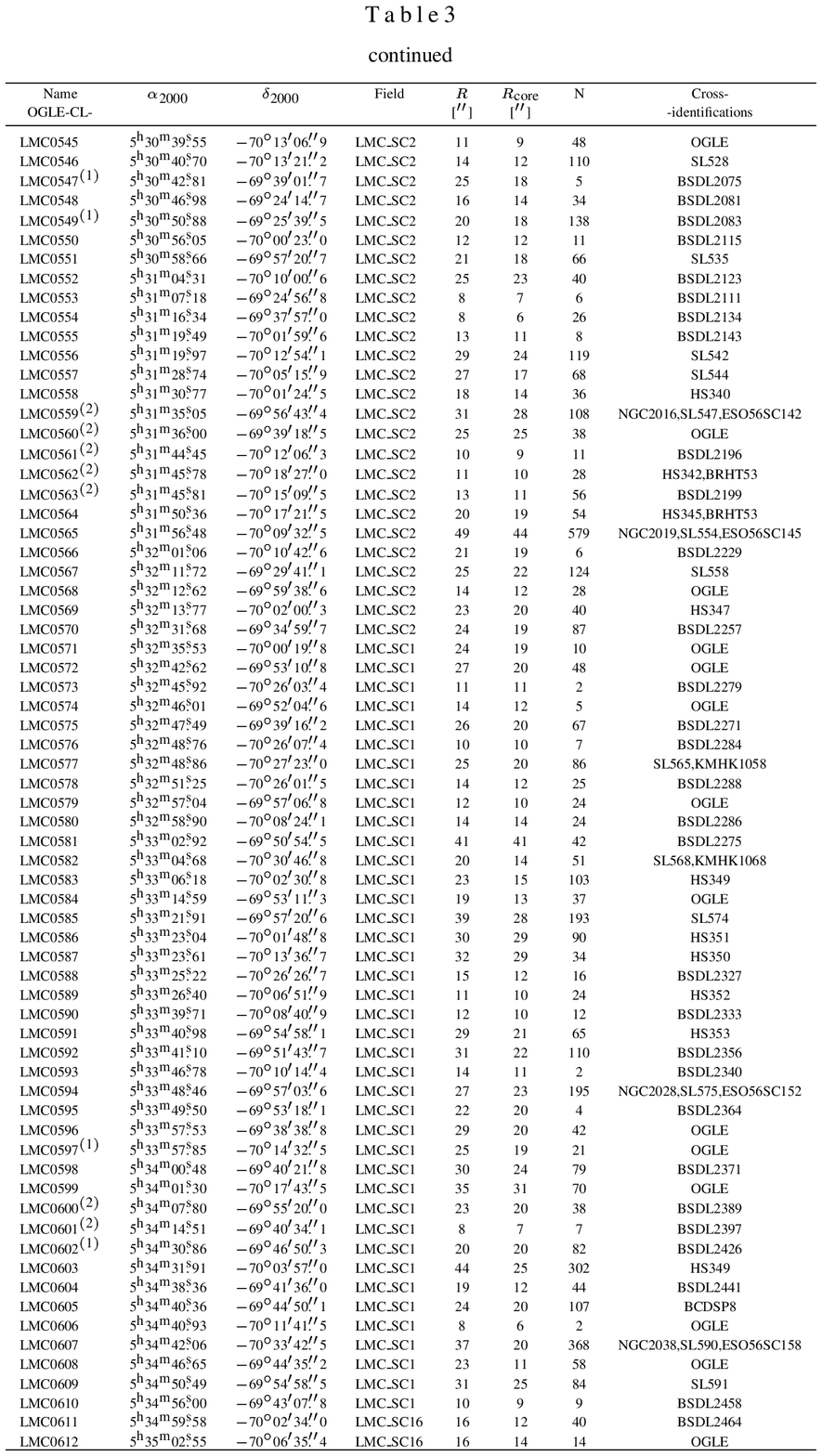,clip=}} \end{figure}
\begin{figure}[p] \vspace*{-6.3cm}
\centerline{\hspace*{10mm}\psfig{figure=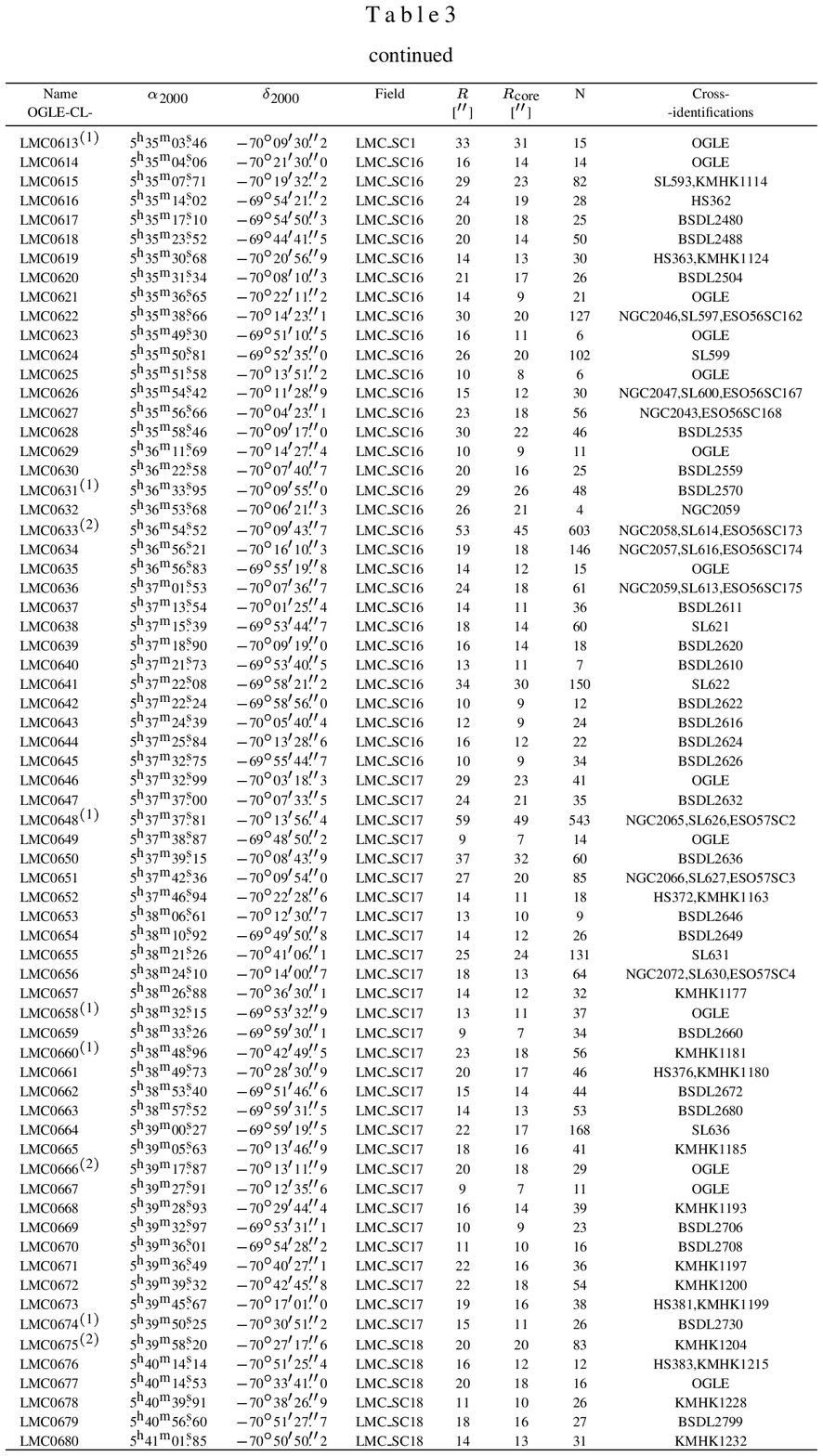,clip=}} \end{figure}
\begin{figure}[p] \vspace*{-6.3cm}
\centerline{\hspace*{10mm}\psfig{figure=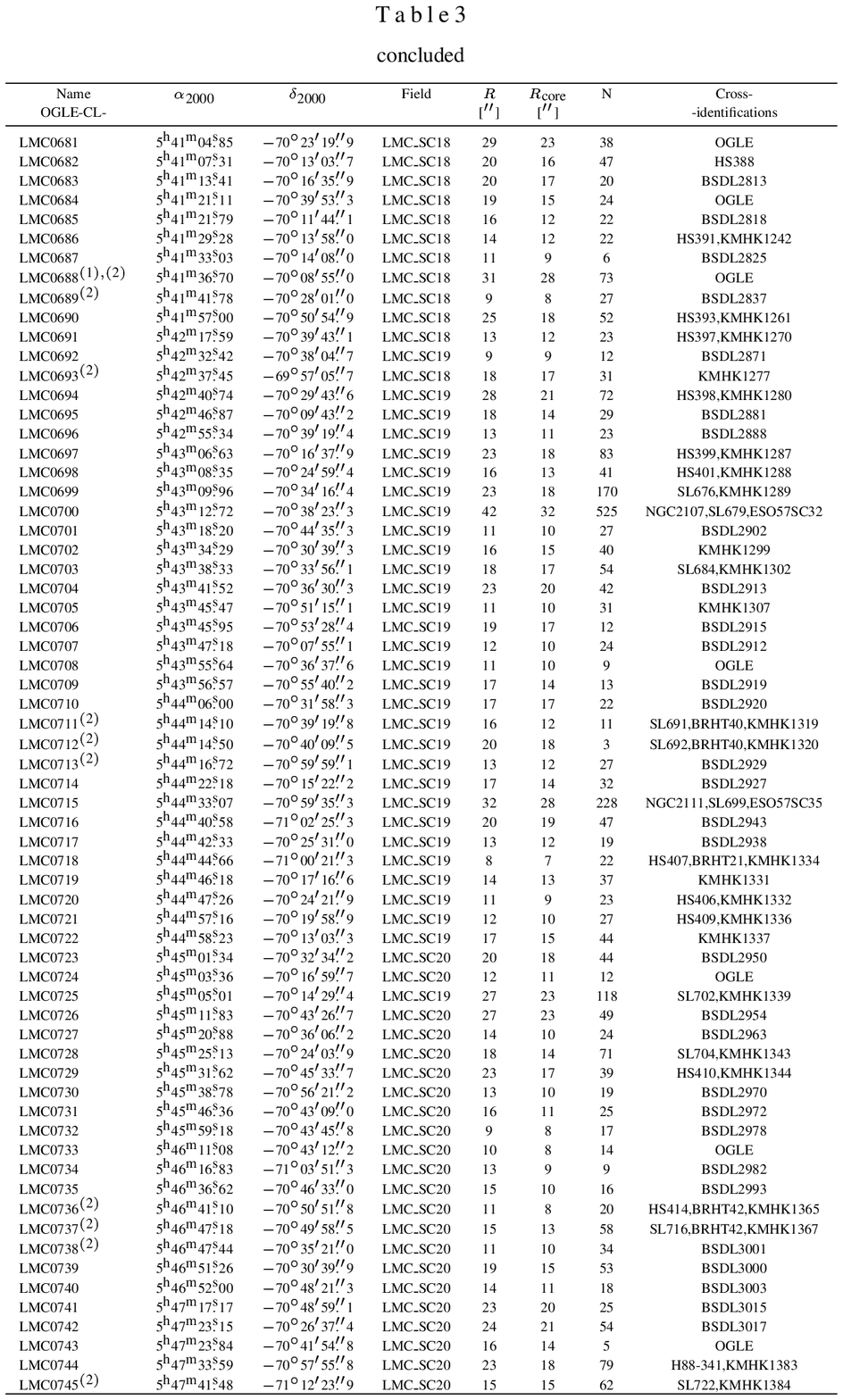,clip=}} \end{figure}
\Subsection{Content of the Catalog} Table~3 presents  the OGLE Catalog
of star clusters in the LMC. Information on  all 745 clusters detected
in this project is given. Column 1 contains the OGLE  identification of
the cluster, consisting of the prefix OGLE-CL-LMC and four  digit
number. In columns 2 and 3 we list the equatorial coordinates of  
cluster center and in column 4 the OGLE name of the field in which a
given  cluster was detected. Columns 5, 6, 7 and 8 provide radius, core
radius, crude  number of cluster members and cross-identification (see
Table~2 for  explanation of acronyms used in column 8). Remarks given in
column~1 have the  following meaning : (1)~-- cluster contains bright
star, (2)~-- object is  located close to the edge of the frame or bad
columns. Crude number of members was calculated by subtraction of the
mean stellar background from the number of stars counted in the
radius of the cluster. Stellar background was averaged from independent
counts in four regions around the cluster. While for the more populous
clusters this number quite reasonably approximate number of stars in the
cluster, its meaning is "several" when it is below a dozen or so
(typically for small clusters located in high stellar background regions).

Appendix presents a few pages of the Atlas of star clusters  from the
LMC. It  consists of the finding chart and color-magnitude diagrams
(CMD) of each  cluster. Presented CMDs were not cleaned for field stars
so in the case of  clusters located in dense stellar regions and objects
with small number of stars they can be contaminated by field stars. One
can perform subtraction of field stars from  cluster CMDs when the OGLE
photometric maps of the LMC are released (Udalski  \etal in
preparation). Full version of Atlas is available electronically from 
the OGLE Internet archive. 

\Section{Summary}
We present the catalog of clusters found in the 5.8 square degrees area of the 
LMC, based on the {\it BVI} observations collected in the course of the  
OGLE-II microlensing project. The automatic, algorithmic procedure similar to 
that used in searching for clusters in the SMC (Paper I) resulted in 
detection of 745 objects. 126 of them are the new ones. For all of them the 
equatorial coordinates, radius, approximate number  of members and 
cross-identification with previous catalogs are provided. 

The Catalog, full version of the Atlas and {\it BVI} photometry of each 
cluster can be obtained from the OGLE Internet archive: \newline 
\centerline{{\it ftp://sirius.astrouw.edu.pl/ogle/ogle2/clusters/lmc/} or } 
\centerline{{\it http://www.astrouw.edu.pl/\~{}ogle}}

\noindent
and its US mirror 

\centerline{{\it http://www.princeton.edu/\~{}ogle}.}

\Acknow{ We would like to thank Dr.\ Eduardo Bica for kindly providing us with 
the computer version of the list of clusters from the LMC. The  paper was 
partly supported by the Polish KBN grants: 2P03D00814 to A.\ Udalski and 
2P03D00617 to G.\ Pietrzy{\'n}ski. Partial support for the OGLE project was 
provided with the NSF grant  AST-9820314 to B.~Paczy\'nski.}

\begin{figure}[p]
\vspace*{-5.1cm}
\hglue5.5cm{\psfig{figure=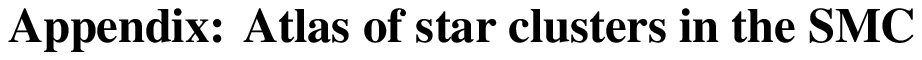,bbllx=250pt,bblly=815pt,bburx=400pt,bbury=842pt,angle=0}}
\vskip1cm
\hglue-4cm{\psfig{figure=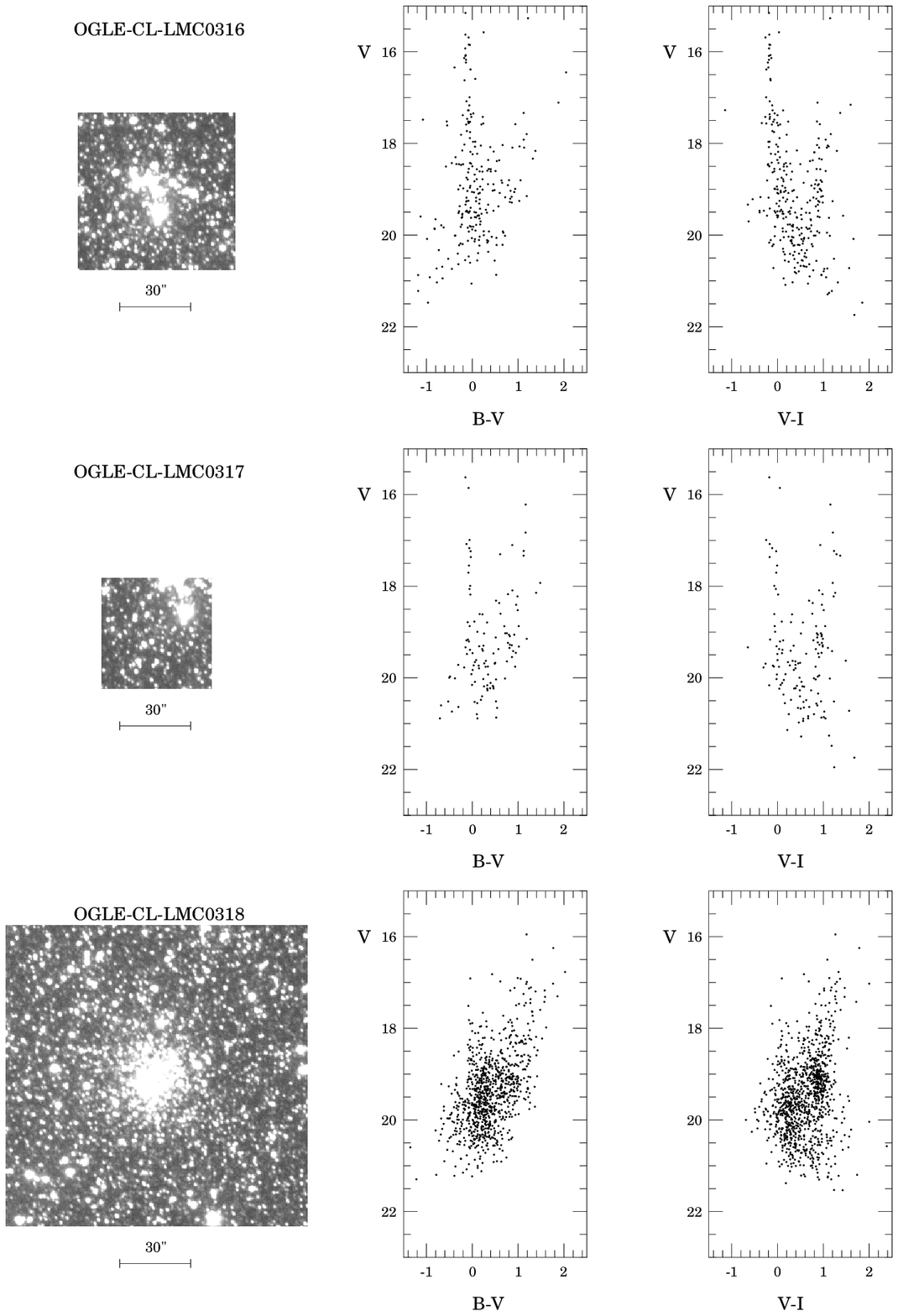,angle=0}}
\end{figure}
\end{document}